\documentclass[12pt,a4paper]{article}
\usepackage[utf8]{inputenc}
\usepackage{amsmath}
\usepackage{amssymb}
\usepackage{amsthm}
\usepackage{multirow}
\usepackage{bm}
\usepackage{float}
\usepackage{amsfonts}
\usepackage{graphicx}
\usepackage[]{cite}
\usepackage{float}
\usepackage{verbatim}
\usepackage[left=2cm,right=2cm,top=3cm,bottom=2.5cm]{geometry}

\begin{document}
\begin{center}
\textbf{Observational Constraints and Cosmic Growth Index of Realistic $f(G)$ Gravity Frameworks using MCMC Analysis}
\end{center}
\hfill\\
Praveen Kumar Dhankar$^{1}$, Albert Munyeshyaka$^{2}$, Mohit Thakre$^{1}$, Joseph Ntahompagaze$^{3}$, Safiqul Islam$^{4}$ and Farook Rahaman$^{5}$\\
\hfill\\ 
$^{1}$Symbiosis Institute of Technology, Nagpur Campus, Symbiosis International (Deemed University), Pune, India\;\;\; \; \;\hfill\\
$^{2}$Department of Education, East African University Rwanda, Nyagatare, Rwanda\;\;\; \; \;\hfill\\ 
$^{3}$Department of Physics, College of Science and Technology, University of Rwanda,Kigali, Rwanda\;\;\; \; \;\hfill\\ 
$^{4}$ Department of Mathematics and Statistics, College of Science, King Faisal University, Al Ahsa, Saudi Arabia
\;\;\; \; \;\hfill\\
$^{5}$ Department of Mathematics, Jadavpur University, Kolkata, India\\ \\\\
Correspondence:munalph@gmail.com\;\;\;\;\;\;\;\;\;\;\;\;\;\;\;\;\;\;\;\;\;\;\;\;\;\;\;\;\;\;\;\;\;\;\;\;\;\;\;\;\;\;\;\;\;\;\;\;\;\;\;\;\;\;\;\;\;\;\;\;\;\;\;\;\;\;\;\;\;\;\;\;\;\;\;
\begin{center}
\textbf{Abstract}
\end{center}
We present a comprehensive observational analysis of modified Gauss--Bonnet, or $f(G)$, gravity by investigating both the background cosmological expansion and sub-horizon linear matter perturbations. We consider two viable functional forms: an arctangent parameterization (Model~I) and a generalized polynomial power-law model (Model~II). Using a Markov Chain Monte Carlo (MCMC) ensemble sampler, we constrain their parameter spaces through background and structure-growth observations. Our analysis employs cumulative combinations of Cosmic Chronometers (CC), the Pantheon+ Type Ia Supernovae compilation (PP), Redshift-Space Distortions (RSD), and the Year-1 Baryon Acoustic Oscillation measurements from the Dark Energy Spectroscopic Instrument (DESI BAO). The inclusion of DESI BAO data produces a noticeable downward shift in the preferred values of the Hubble constant, $H_0$, and present matter density parameter, $\Omega_{m0}$. For Model~II, the full analysis yields $H_0=64.02^{+4.13}_{-3.08},\mathrm{km,s^{-1},Mpc^{-1}}$ and $\Omega_{m0}=0.249^{+0.047}_{-0.038}$. We further assess the statistical performance of the models relative to $\Lambda$CDM using $\Delta\mathrm{AIC}_c$, $\Delta\mathrm{BIC}$, and $\Delta\mathrm{DIC}$. For Model~II, the full dataset gives $\Delta\mathrm{AIC}_c=4.233$ and $\Delta\mathrm{DIC}=5.971$, favoring the $\Lambda$CDM baseline. At the perturbation level, both models exhibit stable growth histories compatible with large-scale structure observations. The models predict transitions in the late-time expansion dynamics at $z\approx0.5005$ and $z\approx0.6086$ for Models~I and II, respectively, highlighting differences in their cosmological evolution.
\\
\hfill\\
\textit{keywords:} Cosmic acceleration--covariant formalism--large scale structure-- modified gravity--MCMC analysis-- Observational constraints.
\section{Introduction}\label{introduction}
The remarkable success of General Relativity (GR) in explaining gravitational phenomena across a wide range of scales has established it as the standard theory of gravitation. Classical tests such as the deflection of light by the Sun, the perihelion advance of Mercury, gravitational redshift, and the recent direct detection of gravitational waves have provided strong confirmation of its predictions \cite{dyson1920ix,janssen2021einstein,abbott2016ligo,landau2013classical,ishak2019testing,hazarika2024f}. Despite these achievements, several observational and theoretical challenges indicate that GR may not represent the ultimate description of gravity. In particular, the theory encounters difficulties when applied to quantum scales and fails to naturally explain the observed accelerated expansion of the Universe without introducing additional components.

The discovery of cosmic acceleration through observations of Type Ia supernovae, cosmic microwave background anisotropies, and large-scale structure surveys \cite{riess1998observational,perlmutter1999measurements,tonry2003cosmological,spergel2007three,aghanim2021erratum} have motivated extensive efforts to understand the underlying mechanism responsible for this phenomenon. Two broad approaches have emerged. The first assumes the existence of an exotic dark-energy component with negative pressure, such as scalar fields, Chaplygin gas models, or a cosmological constant \cite{copeland2006dynamics,saadat2013viscous,sahlu2019chaplygin,gadbail2022generalized,sahlu2023confronting,carroll2001cosmological}. The second approach seeks modifications of the gravitational sector itself, leading to a variety of extended theories of gravity.

Among the most studied alternatives are $f(R)$, $f(T)$, $f(Q)$, and $f(G)$ gravity theories\cite{nashed2024constraining,escamilla2024f,pawar2024two,gadbail2024modified,maurya2024modified,makarenko2017asymptotic,nojiri2007introduction,munyeshyaka2024covariant,Red2}, where $R$, $T$, $Q$, and $G$ denote the Ricci scalar, torsion scalar, nonmetricity scalar, and Gauss--Bonnet invariant, respectively. These modified theories offer the possibility of explaining both the inflationary epoch and the present accelerated expansion through purely geometric effects. Nevertheless, any viable modified gravity theory must satisfy both theoretical consistency conditions and observational constraints.

Modified Gauss--Bonnet gravity, in particular, has attracted considerable attention due to its ability to produce rich cosmological dynamics while avoiding some of the shortcomings associated with simpler extensions of GR. Several studies have shown that suitable choices of the function $f(G)$ can successfully reproduce the transition from decelerated to accelerated expansion and can generate viable cosmological histories \cite{makarenko2017asymptotic,nojiri2005modified,lee2020viable}. However, higher-order curvature corrections may also introduce instabilities, including ghost and gradient modes, which necessitate careful theoretical and observational examination \cite{hikmawan2016comment,charmousis2008instability,odintsov2023inflation}. Although a complete stability analysis is beyond the scope of the present work, observational constraints can provide valuable information regarding the physically admissible parameter space of these models.

Determining observationally viable forms of $f(G)$ remains a central challenge in modified gravity research. While theoretical requirements such as stability conditions, symmetry considerations, and consistency with local gravity tests can guide model construction, confrontation with cosmological observations remains indispensable \cite{anagnostopoulos2019bayesian}. Consequently, datasets such as Type Ia supernovae, cosmic chronometers, baryon acoustic oscillations, redshift-space distortions, and cosmic microwave background measurements have become powerful tools for constraining alternative cosmological scenarios \cite{capozziello2015transition,bonici2019constraints,pan2024interacting,anagnostopoulos2019bayesian}.

Recent observations from the Dark Energy Spectroscopic Instrument (DESI) have further improved the precision of cosmological parameter estimation. Several studies have demonstrated that DESI BAO measurements significantly tighten constraints on dark-energy and modified-gravity models while helping to reduce parameter degeneracies \cite{wang2024constraining,pang2024reevaluating,zheng2024cosmological,luciano2025barrow}. Such developments make it timely to reassess the viability of modified Gauss--Bonnet cosmologies using the latest observational data.

A number of recent investigations have focused on both theoretical and observational aspects of $f(G)$ gravity. Cosmological applications ranging from bouncing solutions and late-time acceleration to data-driven reconstructions have been explored extensively \cite{makarenko2017asymptotic,nojiri2005modified,lee2020viable}. In addition, several observational studies have constrained different classes of $f(G)$ models using combinations of supernova, Hubble parameter, BAO, and large-scale structure datasets \cite{dhankar2025constraints,dhankar2026testing,dhankar2026constraints,munyeshyaka2026constraining,DHANKAR2026140485,munyeshyaka2026background,munyeshyaka2021cosmological,munyeshyaka2025matter,munyeshyaka20231+,munyeshyaka2023multifluid,munyeshyaka2023perturbations,dhankar2026large}. These analyses indicate that modified Gauss--Bonnet gravity remains a promising framework for describing the late-time evolution of the Universe beyond the standard $\Lambda$CDM paradigm.

Motivated by the above developments, the present work aims to constrain cosmological parameters within the framework of modified Gauss--Bonnet gravity at both the background and perturbation levels. This work focusses on two representative and phenomenologically interesting functional forms of $f(G)$. The first model is given by $f(G)=\lambda \frac{G}{G_{0}}\arctan\left(\frac{G}{G_{0}}\right)-\alpha\lambda\sqrt{G_{0}}$, \cite{DeFelice2009,Nojiri2005,de2009construction}
which will be referred to as Model I. The second model is defined as
$f(G)=G^{m}\left(a+bG^{n}\right)$ \cite{de2009construction}, and will be referred to as Model II. Here, $\lambda$, $\alpha$, $G_{0}$, $a$, $b$, $m$, and $n$ are free parameters to be constrained by observational data. 
Starting from the modified Friedmann equations associated with these two choices, our primary objective is to place observational constraints on the corresponding parameter spaces through a Bayesian framework. To achieve this goal, one employs recent cosmological observations including cosmic chronometer Hubble measurements (CC), the Pantheon Plus Type Ia supernova compilation (PP), redshift-space distortion measurements (RSD), and DESI baryon acoustic oscillation (BAO) data. Throughout the analysis, the cosmic matter content is assumed to consist of pressure-less dust with equation-of-state parameter $w=0$.

In addition to the background evolution, we investigate the growth of matter density perturbations. The evolution of the matter density contrast $\delta_m$ is governed by
$\ddot{\delta}_{m}+2H\dot{\delta}_{m}-4\pi G_{\rm eff}\rho_{m}\delta_{m}=0$,
where $G_{\rm eff}$ denotes the effective gravitational coupling predicted by the underlying $f(G)$ model. This equation allows  to study the impact of Gauss--Bonnet corrections on the growth rate of cosmic structures and to confront the theoretical predictions with redshift-space distortion observations.

The modified background and perturbation equations are solved numerically without invoking analytical approximations. To obtain reliable bounds on the free model parameters, one performs a Markov Chain Monte Carlo (MCMC) analysis and derive the corresponding posterior probability distributions and corner plots. Parameter estimation is carried out using the individual datasets PP, CC, RSD, and DESI BAO, as well as their combinations PP+CC, PP+CC+RSD, and PP+CC+RSD+DESI BAO. This approach enables  to assess the constraining power of each observational probe and the impact of dataset complementarity.

Furthermore, we perform a statistical comparison between the two modified Gauss--Bonnet models and the standard $\Lambda$CDM cosmology using the Akaike Information Criterion (AIC), Bayesian Information Criterion (BIC), and Deviance Information Criterion (DIC). Finally, employing the best-fit parameter values reconstructs the cosmological evolution of the Universe through the deceleration parameter $q$ and the growth of matter perturbations, thereby assessing whether the considered $f(G)$ models can simultaneously account for the observed late-time acceleration and the formation of large-scale structures.
 Furthermore, one compare the statistical performance of the proposed models with that of the standard $\Lambda$CDM cosmology through the Akaike Information Criterion (AIC), Bayesian Information Criterion (BIC), and Deviance Information Criterion (DIC) and finally analyse the reconstructed deceleration parameter and related cosmological quantities to assess the ability of the considered $f(G)$ models to account for the observed late-time acceleration of the Universe.

Although numerous investigations of modified Gauss--Bonnet gravity have appeared in the literature, comprehensive analyses that simultaneously confront different viable $f(G)$ functional forms with the latest Pantheon Plus, cosmic chronometer, redshift-space distortion, and DESI BAO datasets remain relatively scarce. The present work therefore provides a unified Bayesian comparison of two distinct Gauss--Bonnet models within the same numerical and statistical framework. Such an approach allows a direct assessment of their observational viability and offers new insights into whether geometrically modified gravity can provide a realistic alternative to the concordance $\Lambda$CDM cosmology.

The remaining part of this paper is organized as follows. Sec.~\ref{ba},  presents the field equations of modified Gauss--Bonnet gravity and derive the cosmological equations corresponding to the two models under consideration.  Sec.~\ref{da} describes the observational datasets and the Bayesian methodology employed in the analysis. The parameter constraints and statistical model comparison are presented in Sec.~\ref{re}. This section also discusses the cosmological implications of the best-fit solutions through the evolution of the deceleration parameter and other relevant observables. Finally, concluding remarks are given in Sec.~\ref{con}.

\section{Cosmological Framework and Background Dynamics in Modified Gauss-Bonnet Gravity}\label{ba}

To explore the late-time acceleration of the cosmos, one considers an extended gravitational action where the standard Einstein-Hilbert term is modified by a generic function of the Gauss-Bonnet invariant, $f(G)$. Incorporating a minimally coupled matter sector, the total action is formulated as \cite{nojiri2011unified,li2007cosmology,venikoudis2022late}:
\begin{eqnarray}
 S=\int d^{4}x\sqrt{-g}\Big(\frac{R}{2\kappa^{2}}+\frac{f(G)}{2}+\mathcal{L}_{m}\Big)\;,
 \label{eq1}
\end{eqnarray}
where $g$ denotes the metric determinant, $R$ is the Ricci scalar, $\kappa^2 = 8\pi G_N$ represents the gravitational coupling constant, and $\mathcal{L}_{m}$ represents the Lagrangian density of cosmic matter fields. Notably, if $f(G) = G$, the integral contribution $\int d^{4}x\sqrt{-g}G$ acts as a topological invariant in four dimensions, meaning its variation vanishes identically ($\delta \int d^4x \sqrt{-g}G = 0$). Consequently, the action seamlessly reduces to that of standard General Relativity (GR). Varying Eq.~(\ref{eq1}) with respect to the metric tensor yields the effective gravitational field equations:
\begin{eqnarray}
 G_{ij} \equiv R_{ij}-\frac{1}{2}g_{ij}R=8\pi G_{N}T^{tot}_{ij}\;,\label{eq2}
\end{eqnarray}
where $G_{ij}$ is the Einstein tensor and $T^{tot}_{ij}$ encapsulates the cumulative energy-momentum contributions from all physical fluids. Modeling the cosmic substrate as a perfect fluid, its constituent energy-momentum tensor takes the standard form:
\begin{eqnarray}
 T_{ij}=(\rho+p)u^{i}u^{j}+pg_{ij}\;,
\end{eqnarray}
with $\rho$, $p$, and $u^{i}$ designating the energy density, isotropic pressure, and the respective comoving 4-velocity vector.

Assuming a spatially flat, homogeneous, and isotropic universe, we adopt the standard Friedmann-Robertson-Walker (FRW) line element:
\begin{eqnarray}
ds^{2}=-dt^{2}+a(t)^{2}\Big(dx^{2}+dy^{2}+dz^{2}\Big)\;,
\end{eqnarray} 
where $a(t)$ denotes the cosmic scale factor. Under this geometric configuration, the Ricci scalar $R$ and the Gauss-Bonnet invariant $G$ evolve according to the Hubble parameter $H \equiv \frac{\dot{a}}{a}$ via:
\begin{equation}
R=6\Big(\dot{H}+2H^{2}\Big)\;, \quad \text{and} \quad G=24H^{2}\Big(\dot{H}+H^{2}\Big)\;.
\end{equation}
By substituting these invariants into the field equations, the modified Friedmann and Raychaudhuri equations are obtained as:
\begin{eqnarray}
&& 3H^{2}=\rho_{m}+\rho_{r}+\rho_{G}\;,\label{eq2.5}\\
&&\Big(2\dot{H}+3H^{2}\Big)=-p_{m}-p_{r}-p_{G}\;,
\end{eqnarray}
where the effective energy density $\rho_{G}$ and pressure $p_{G}$ attributed to the geometric Gauss-Bonnet fluid are explicitly given by:
\begin{eqnarray}
&&\rho_{G}=-\frac{1}{2}f(G)+\frac{1}{2}Gf'(G)\;,\label{eq2.7}\\
&&p_{G}=-\frac{1}{2}Gf'(G)+\frac{1}{2}f(G)-12H^{3}\dot{f}'(G)\;.
\end{eqnarray} 
Here, primes denote differentiation with respect to $G$. The energy densities for non-relativistic dark matter ($\rho_{m}$) and radiation ($\rho_{r}$) scale with redshift $z$ according to $\rho_{m}=\rho_{m0}(1+z)^{3}$ and $\rho_{r}=\rho_{r0}(1+z)^{4}$, while $p_m$ and $p_G$ denote their respective isotropic pressures. Assuming a barotropic equation of state $p_i = w_i \rho_i$ for each distinct component, local energy conservation leads to isolated continuity equations:
\begin{eqnarray}
&& \dot{\rho}_{m}+\theta \Big(\rho_{m}+p_{m}\Big)=0\;,\\
&&\dot{\rho}_{G}+\theta \Big(\rho_{G}+p_{G}\Big)=0\;,
\end{eqnarray}
where $\theta = 3H$ defines the fluid expansion scalar. 

To facilitate comparison with observational data, we introduce the normalized, dimensionless Hubble parameter $E^2 \equiv \frac{H^{2}}{H^{2}_{0}}$. Utilizing the modified Friedmann constraints from Eqs.~(\ref{eq2.5}) and (\ref{eq2.7}), the cosmic expansion rate can be re-expressed as:
\begin{equation}
E^{2}=\Omega_{m0}\Big(1+z\Big)^{3}+\Omega_{r0}\Big(1+z\Big)^{4}+\Omega_{f0}y(z,r)\;,\label{eq2.11}
\end{equation}
where the respective current-day density parameters satisfy the normalization condition $\Omega_{f0} = 1 - \Omega_{m0} - \Omega_{r0}$. From this framework, it directly follows that the geometric correction term $y(z,r)$ can be isolated as:
\begin{equation}
y(z,r)=\frac{1}{6H^{2}_{0}\Omega_{f0}}\Big[Gf'(G)-f(G)\Big]\;.\label{eq2.12}
\end{equation}
The explicit functional evolution of $y(z,r)$ depends entirely on the choice of the $f(G)$ gravity models, which will be explored systematically in the subsequent sections.
\subsection{Explicit Formulations of $f(G)$ Gravity Models}
In this section, we investigate two distinct functional profiles within the framework of modified Gauss-Bonnet gravity. Specifically, we present an arctangent-based parameterization and a generalized polynomial power-law model. The background cosmological evolution of these formulations is subsequently examined using a suite of modern observational datasets.
\subsubsection{Model I: Arctangent $f(G)$ Parameterization}
We first consider a realistic modified Gauss-Bonnet profile designed to smoothly mimic late-time cosmic acceleration while bypassing solar system constraints. The functional form is given by \cite{nojiri2005modified,lee2020viable,DeFelice2009,Nojiri2005,de2009construction}: 
\begin{equation}
f(G)= \lambda \frac{G}{G_{0}}\arctan\left(\frac{G}{G_{0}}\right)-\alpha \lambda \sqrt{G_{0}} \label{eq2.13}\;,
\end{equation}
where $\alpha$, $\lambda$, and $G_{0}$ represent the characteristic scale parameters of the model. Here, $G_0$ is conventionally scaled to the current epoch value $G_{0}=-24q_{0}H^{4}_{0}$, where $q_{0}$ and $H_{0}$ are the present deceleration and Hubble parameters. Taking the first derivative of Eq.~(\ref{eq2.13}) with respect to $G$ yields:
\begin{equation}
f'(G) = \frac{\lambda}{G_0}\arctan\left(\frac{G}{G_0}\right) + \frac{\lambda G}{G^2 + G_0^2}\;.
\end{equation}
Substituting this result into the geometric correction definition in Eq.~(\ref{eq2.12}), the dimensionless function $y(z,r)$ evaluates to:
\begin{equation}
y(z,r)= \frac{\lambda}{6H^{2}_{0}\Omega_{f0}}\left[\frac{G^2}{G^2 + G_0^2} + \alpha\sqrt{G_0}\right]\;.
\label{eq:y_model1}
\end{equation}
By substituting Eq.~(\ref{eq:y_model1}) back into the generalized Friedmann constraint, the explicit cosmic expansion rate for Model I is mapped as:
\begin{equation}
H^2 = H_0^{2}\Omega_{m0}\Big(1+z\Big)^{3} + H_0^2\Omega_{r0}\Big(1+z\Big)^{4} + \frac{\lambda}{6}\left[\frac{G^2}{G^2 + G_0^2} + \alpha\sqrt{G_0}\right]\;.
\label{F11}
\end{equation}
Equation~(\ref{F11}) establishes the background hubble evolution equation governing the arctangent modified gravity sector.

\subsubsection{Model II: Generalized Polynomial $f(G)$ Parameterization}
As an alternative extension motivated by non-linear curvature modifications frequently implemented in $f(R)$ and $f(T)$ theories \cite{linder2009exponential,nesseris2013viable}, we introduce a generalized polynomial power-law $f(G)$ framework defined as \cite{de2009construction}:
\begin{equation}
f(G)=  G^{m}\Big(a+bG^{n}\Big) \label{eq2.16}\;,
\end{equation}
where $a$, $b$, $m$, and $n$ denote arbitrary dimensionless constants. Employing an identical analytical procedure via Eq.~(\ref{eq2.12}), the corresponding geometric contribution simplifies to:
\begin{equation}
y(z,r)= \frac{1}{6H^{2}_{0}\Omega_{f0}}\Big[a(m-1)+b(m+n-1)G^{n}\Big]G^{m}\;,
\end{equation} 
which leads directly to the modified background Friedmann expression:
\begin{eqnarray}
&&H^2= H_0^{2} \Omega_{m0}(1+z)^{3} + H_0^2\Omega_{r0}\Big(1+z\Big)^{4} \nonumber\\&&+ \frac{1}{6}\Big[a(m-1)+b(m+n-1)G^{n}\Big]G^{m}\;.
\label{F22}
\end{eqnarray}
Equations~(\ref{F11}) and (\ref{F22}) constitute the core equations utilized to reconstruct the background dynamics of the universe.

\subsubsection{Physical Motivation of the Considered $f(G)$ Models}

The selection of viable $f(G)$ functions is motivated by both theoretical consistency and observational requirements. In particular, a realistic modified Gauss--Bonnet gravity model should: (i) reproduce the observed late-time accelerated expansion of the Universe without the need for an explicit cosmological constant, (ii) recover General Relativity in high-curvature regimes to satisfy local gravity constraints, and (iii) remain free from pathological instabilities. The two models considered in this work are motivated by these criteria from complementary perspectives.

\subsubsection{Motivation for Model I: Arctangent $f(G)$ Parameterization}

The arctangent model is designed to provide a smooth and bounded modification of gravity that becomes significant only at low curvature scales associated with the recent cosmological epoch. The functional form
\begin{equation}
f(G)=\lambda \frac{G}{G_0}\arctan\left(\frac{G}{G_0}\right)-\alpha \lambda \sqrt{G_0},
\end{equation}
possesses several attractive physical properties.

First, the model naturally recovers General Relativity at high curvatures. In the limit $G\gg G_0$, the arctangent function approaches a constant value,
\begin{equation}
\arctan\left(\frac{G}{G_0}\right)\rightarrow \frac{\pi}{2},
\end{equation}
which suppresses large deviations from standard cosmology during the radiation- and matter-dominated eras. Consequently, the model remains compatible with solar-system and astrophysical observations.

Second, around the present cosmological epoch where $G\sim G_0$, the non-linear Gauss--Bonnet correction becomes dynamically relevant and effectively behaves as a dark-energy component capable of driving the observed late-time acceleration without introducing an additional cosmological constant.

Third, the arctangent function is smooth and finite for all values of $G$, avoiding singular behaviors that may arise in simpler power-law parameterizations. This property contributes to the theoretical stability and numerical robustness of the model.

Finally, because the correction term saturates at large curvature, the model naturally suppresses modified-gravity effects in high-density environments, providing an effective screening mechanism that helps satisfy local gravity constraints.

\subsubsection{Motivation for Model II: Generalized Polynomial $f(G)$ Parameterization}

The generalized polynomial model
\begin{equation}
f(G)=G^{m}\left(a+bG^{n}\right),
\end{equation}
is motivated by the broad class of curvature-based modifications frequently employed in alternative theories of gravity, including $f(R)$, $f(T)$, and higher-order curvature models.

One of its principal advantages is its flexibility in describing different cosmological regimes. The free exponents $m$ and $n$ determine how the Gauss--Bonnet contribution scales with curvature, allowing the model to reproduce a wide range of cosmic histories, including matter domination, quintessence-like behavior, and accelerated expansion.

Furthermore, polynomial curvature corrections naturally emerge in effective gravitational actions inspired by quantum gravity, string theory, and higher-dimensional scenarios. The model can therefore be regarded as a phenomenological representation of possible higher-order geometric corrections to Einstein's theory.

Another attractive feature is its potential to provide a unified description of the early and late Universe. Depending on the choice of the parameters $m$ and $n$, the Gauss--Bonnet correction may become important either at high curvature, relevant to inflationary dynamics, or at low curvature, relevant to dark-energy phenomena.

In addition, the analytical simplicity of the polynomial structure facilitates both theoretical investigations and observational parameter estimation. Several previously proposed viable $f(G)$ models can be recovered as special cases through suitable choices of the parameters $a$, $b$, $m$, and $n$, making the framework sufficiently general for comprehensive cosmological analyses.

In summary, the arctangent parameterization is primarily motivated by the construction of a viable late-time accelerating cosmology with suppressed high-curvature deviations from General Relativity, whereas the generalized polynomial model is motivated by its connection to higher-order curvature corrections and its flexibility in describing different cosmological epochs. Together, these models provide complementary frameworks for investigating the role of Gauss--Bonnet modifications in the expansion history of the Universe.

To trace the transition from the early deceleration epoch to the ongoing accelerated expansion phase, we monitor the evolution of the deceleration parameter $q \equiv -\frac{\ddot{a}a}{\dot{a}^{2}}$. Expressed in terms of redshift-space differentiation, this diagnostic parameter reads:
\begin{equation}
q(z) = -1 + (1 + z) \frac{1}{H(z)}\frac{dH(z)}{dz}\;.\label{F33}
\end{equation} 
Differentiating the respective Hubble equations with respect to $z$ yields the explicit analytical tracks of the deceleration parameters for Model I ($q_1$) and Model II ($q_2$):
\begin{equation}
\begin{aligned}
    q(z)_{1}&=-1+\frac{(1+z)}{H^2(z)}\Bigg[\frac{3}{2} H_0^2 \Omega_{m0}(1+z)^2 + 2H_0^2\Omega_{r0}(1+z)^3 \\&+ \frac{\lambda G_0^2 G}{3(G^2+G_0^2)^2}\frac{dG}{dz}\Bigg]
\end{aligned}
\end{equation}

\begin{equation}
\begin{aligned}
   q(z)_{2}&=-1+\frac{(1+z)}{H^2(z)}\Bigg[\frac{3}{2}H_0^2 \Omega_{m0}(1+z)^2 + 2H_0^2\Omega_{r0}(1+z)^3 \\&+ \frac{1}{12} \Big[am(m-1)G^{m-1}+b(m+n)(m+n-1)G^{m+n-1}\Big]\frac{dG}{dz} \Bigg]
\end{aligned}
\end{equation}

\begin{figure}[H]
    \centering
    \includegraphics[width=0.5\textwidth]{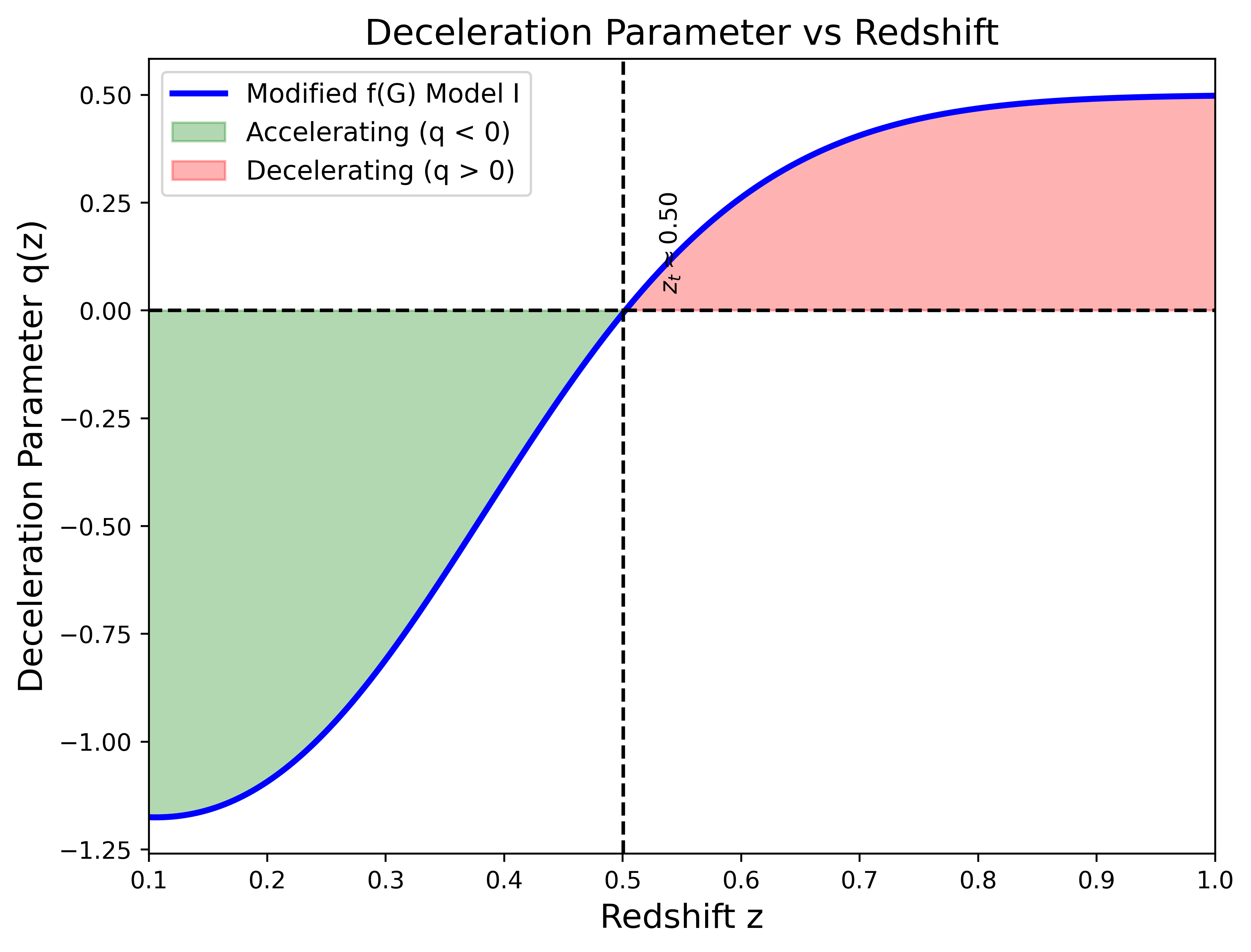} 
    \caption{Evolutionary trajectory of the deceleration parameter $q(z)$ as a function of redshift for Model I.}
     \label{fig:1}
\end{figure}

\begin{figure}[H]
    \centering
    \includegraphics[width=0.5\textwidth]{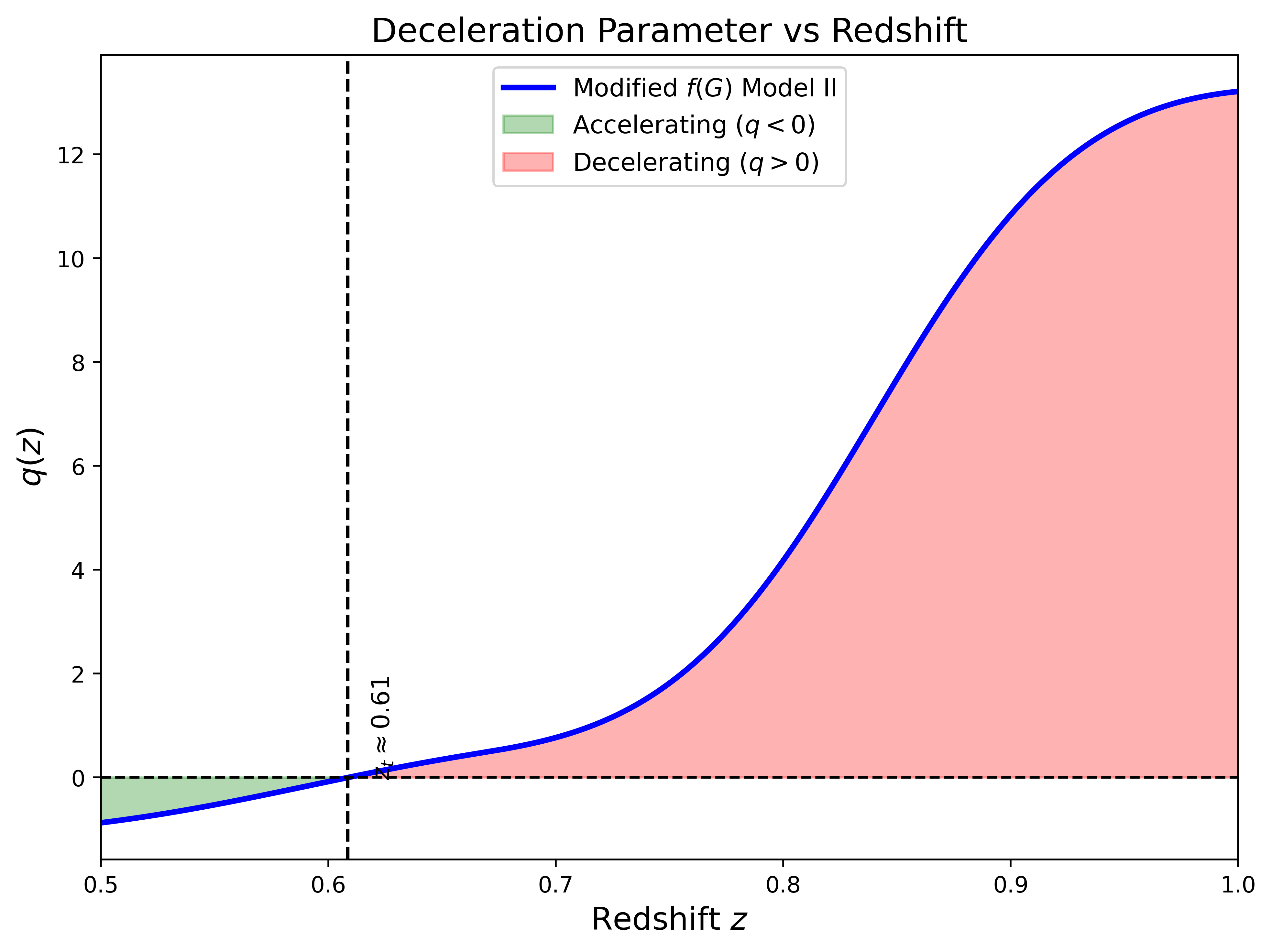} 
    \caption{Evolutionary trajectory of the deceleration parameter $q(z)$ as a function of redshift for Model II.}
     \label{fig:2}
\end{figure}

By integrating Eq.~(\ref{F33}) numerically alongside the constraints provided in Figs.~\ref{fig:1} and \ref{fig:2}, we can robustly map out the cosmic transition threshold for each scenario.

The structural viability of these models hinges upon satisfying key physical criteria. Crucially, a robust $f(G)$ configuration must: (i) seamlessly reproduce the sequential transitions from the radiation era through the matter-dominated epoch to the dark energy era; (ii) exhibit strict consistency with historical and late-time observational constraints; (iii) pass stringent solar system boundary tests; and (iv) ensure the stability of cosmic linear perturbations against small-scale pathologies. Because deviations from the standard $\Lambda$CDM baseline are governed smoothly by the selected model parameters, finding tighter constraints on these parameters is paramount. In the subsequent segment of this study, we outline the statistical methodology and specific observational toolkits utilized for parametric optimization.
\subsection{Linear Cosmological Perturbations and Large-Scale Structure Growth}

To assess the structural stability of the proposed $f(G)$ gravity frameworks and analyze their fingerprints on large-scale structure formation, we investigate linear matter density perturbations. We model the cosmic substrate as a barotropic fluid obeying the equation of state $p = w\rho$. Here, $w=0$ corresponds to a pressureless dust-dominated universe ($p_m=0$), which dictates the epoch of structure growth, while $w=1/3$ governs the highly relativistic, high-density radiation era in the early universe.

The anisotropy of the power spectrum on large scales is not only due to the peculiar galactic velocities but also due to the use of an improper fiducial cosmology $H(z)$ inferred in transforming the recorded angles and redshifts into comoving coordinates in order to formulate the correlation function and the corresponding power spectrum. In particular, the comoving distance between a pair of galaxies spaced by an angle $d\theta$ is derived from the Friedmann Robertson Walker (FRW) metric as 
$d\ell_{\perp} = (1+z)D_A(z)\,d\theta $
where $D_A(z)$ is the angular diameter distance at the redshift of the pair. Also the corresponding spacing along the line of sight is
$d\ell_{\parallel} = \frac{c\,dz}{H(z)}$
where $H(z)$ is the true Hubble expansion rate of the true fundamental cosmology. If a fiducial cosmology $H'(z)$ is assumed instead, the corresponding spacing becomes
$
d\ell'_\perp = (1+z)D'_A d\theta = \left(\frac{D'_A}{D_A}\right) d\ell_\perp = \frac{d\ell_\perp}{f_\perp}
$,
$
d\ell'_\parallel = \frac{c\,dz}{H'} = \left(\frac{H}{H'}\right) d\ell_\parallel = \frac{d\ell_\parallel}{f_\parallel} $
where $F \equiv \frac{f_\parallel}{f_\perp}$ is the induced anisotropy due to the use of incorrect fiducial cosmology and has magnitude 
$F = \frac{f_\parallel}{f_\perp} = \left(\frac{H'}{H}\right) \left(\frac{D'_A}{D_A}\right) $
This prompted anisotropy due to the use of improper fiducial cosmology is the Alcock-Paczynski (AP) effect and is degenerate with the RSD anisotropy produced by the galactic peculiar velocities due to the growth of structures. Thus if an $f\sigma'_{8}$ measurement has been retrieved assuming a fiducial $\Lambda$CDM cosmology $H'(z)$, the corresponding $f\sigma_{8}$ obtained with the true cosmology $H(z)$ is approximated as 
$f\sigma_{8}(z) \simeq \frac{H(z) D_A(z)}{H'(z) D'_A(z)} f\sigma'_{8}(z) \equiv q(z, \Omega_{0m}, \Omega_{0m}') f\sigma'_{8}(z)$

In the sub-horizon approximation, where the perturbation wavenumber satisfies $k \gg aH$, the evolution of the non-relativistic matter density contrast $\delta_m \equiv \delta\rho_m / \rho_m$ is governed by the modified fluid equation:
\begin{equation}
 \ddot{\delta}_{m}+2H\dot{\delta}_{m}-4\pi G_{eff}\rho_{m}\delta_{m} = 0\;, \label{eq3.10}
\end{equation}
where overdots denote differentiation with respect to cosmic time $t$. The modification to the gravitational sector is entirely absorbed into the effective Newtonian coupling constant, defined as \cite{DeFelice2009}:
\begin{equation}
G_{eff} = \frac{G_{N}}{1 + 8H^{2}f''(G)\frac{k^{2}}{a^{2}}}\;,
\end{equation}
with $f''(G) \equiv d^2f/dG^2$. To transform Eq.~(\ref{eq3.10}) into observational redshift space ($z$), we apply the operator conversions $\frac{d}{dt} = -(1+z)H\frac{d}{dz}$ and $\frac{d^2}{dt^2} = (1+z)^2H^2\frac{d^2}{dz^2} + (1+z)H^2\left[1 + (1+z)\frac{1}{H}\frac{dH}{dz}\right]\frac{d}{dz}$. This yields the explicit perturbation differential equation:
\begin{eqnarray}
    &&\delta''_{m} + \left[ \frac{1}{H}\frac{dH}{dz} + \frac{1}{1+z} \right] \delta'_{m} \nonumber\\&&- \frac{3}{2}\frac{H_0^2 \Omega_{m0}(1+z)}{H^2} \left( \frac{G_{eff}}{G_N} \right) \delta_{m} = 0\;,
\label{eq:delta_z}
\end{eqnarray}
where primes denote differentiation with respect to $z$.

\subsubsection{Linear Growth Rate and $f\sigma_{8}(z)$ Diagnostic}
To connect theoretical predictions with observations from the Cosmic Microwave Background (CMB) and galaxy redshift surveys, we define the linear growth factor $f(z) \equiv \frac{d\ln \delta_m}{d\ln a} = -(1+z)\frac{\delta'_{m}(z)}{\delta_{m}(z)}$. Utilizing this definition, the second-order differential equation for the density contrast can be recast into a first-order non-linear differential equation tracking the growth rate:
\begin{eqnarray}
&&(1+z)\frac{df}{dz} = f^{2} - \left[ 1 - \frac{1}{2}\frac{d\ln(E^{2})}{d\ln(1+z)} \right] f \nonumber\\&&- \frac{3}{2}\frac{\Omega_{m0}(1+z)^{3}}{E^{2}} \left( \frac{G_{eff}}{G_N} \right)\;,
\label{eq:growth_rate_f}
\end{eqnarray}
where $E^2(z)$ represents the normalized dimensionless Hubble parameter matching the respective $f(G)$ models. 

To eliminate planetary and galaxy bias from our data comparison, we implement the standard composite observable $f\sigma_{8}(z)$, which maps Redshift-Space Distortions (RSD). This statistic links the expansion-induced growth rate with the root-mean-square mass fluctuation amplitude $\sigma_{8}(z)$ via:
\begin{eqnarray}
f\sigma_{8}(z) = f(z) \sigma_{8}(z) = -(1+z)\sigma_{8}(z_{in})\frac{\delta'_{m}(z)}{\delta_{m}(z_{in})}\;,
\end{eqnarray}
where $z_{in}$ defines an arbitrary initial anchoring redshift chosen deep within the matter-dominated era.

\subsubsection{Growth Dynamics in Model I}
For the arctangent $f(G)$ framework specified in Eq.~(\ref{eq2.13}), the structural modification depends on the analytical second derivative of the Lagrangian with respect to the Gauss-Bonnet invariant, which simplifies directly to:
\begin{eqnarray}
    f''_{1}(G) = \frac{2\lambda G_0^2}{(G^2 + G_0^2)^2}\;.
\end{eqnarray}
Substituting $f''_{1}(G)$ into the effective coupling ratio yields the explicit scaling factor for Model I:
\begin{eqnarray}
\left(\frac{G_{eff}}{G_N}\right)_{\text{Model I}} = \left[ 1 + \frac{16\lambda G_0^2 H^2 k^2}{a^2 (G^2 + G_0^2)^2} \right]^{-1}\;.
\end{eqnarray}
The resulting numerical tracking solutions for Eq.~(\ref{eq:growth_rate_f}) under this model are analyzed through Markov Chain Monte Carlo (MCMC) simulations in Sec.~(\ref{re}).

\subsubsection{Growth Dynamics in Model II}
Applying an identical perturbation strategy to the generalized polynomial power-law model introduced in Eq.~(\ref{eq2.16}), the corresponding second-order curvature derivative is given analytically by:
\begin{eqnarray}
    f''_{2}(G) = \Big[am(m-1) + b(m+n)(m+n-1)G^{n}\Big]G^{m-2}\;.
\end{eqnarray}
This explicitly modifies the gravitational growth driver according to:
\begin{eqnarray}
&&\left(\frac{G_{eff}}{G_N}\right)_{\text{Model II}} = \Big[ 1 + \frac{8H^2 k^2}{a^2}\Big[am(m-1) \nonumber\\&&+ b(m+n)(m+n-1)G^{n}\Big]G^{m-2} \Big]^{-1}\;.
\end{eqnarray}
Solving Eq.~(\ref{eq:growth_rate_f}) with this profile yields the growth trajectories for Model II. By comparing these perturbation channels against standard $\Lambda$CDM baselines using $f\sigma_8$ datasets, we isolate which modification yields the highest statistical alignment with observational structures in Sec.~(\ref{re}).

\section{Observational Datasets and Statistical Methodology}\label{da}

\subsection{Cosmological Datasets}
To determine the empirical viability of the two modified $f(G)$ gravity models, we map their theoretical predictions against an array of low- and intermediate-redshift cosmological observations. The parameter vectors under investigation are defined as $\theta_{\text{I}} = \{\Omega_{m0}, \lambda, \alpha, G_{0}, h\}$ for Model~I and $\theta_{\text{II}} = \{\Omega_{m0}, a, b, m, n, h\}$ for Model~II. Here, the normalized Hubble parameter $h$ scales the present-day cosmic expansion rate according to $h = H_0 / (100 \, \mathrm{km\,s^{-1}\,Mpc^{-1}})$. 

We execute our parameter estimation via a Markov Chain Monte Carlo (MCMC) pipeline \cite{mcmc1}, evaluating three complementary combinations of observational data:
\begin{itemize}
    \item \textbf{Dataset~I}: Cosmic Chronometers + Pantheon Plus (CC + PP)
    \item \textbf{Dataset~II}: Cosmic Chronometers + Pantheon Plus + Redshift-Space Distortions (CC + PP + RSD)
    \item \textbf{Dataset~III}: Cosmic Chronometers + Pantheon Plus + Redshift-Space Distortions + DESI Baryon Acoustic Oscillations (CC + PP + RSD + DESI BAO)
\end{itemize}

The individual components comprising these data combinations are described in detail below:

\subsubsection{Pantheon+ Type Ia Supernovae}
We utilize the updated Pantheon+ compilation of Type Ia Supernovae (SNe Ia) \cite{pp}, which aggregates 1701 light curves across 1550 unique SNe Ia spanning the redshift window $0.001 \leq z \leq 2.3$. The goodness-of-fit is assessed using the standard $\chi^2$ statistic:
\begin{align}
    \chi^2_{\text{PP}} = \vec{F}^{\,T} \cdot C^{-1}_{\text{PP}} \cdot \vec{F}\;,
    \label{chi}
\end{align}
where $C_{\text{PP}}$ denotes the comprehensive covariance matrix modeling intertwined systematic and statistical errors. The matrix acts on the residual vector $\vec{F}$, which maps the difference between the observed and theoretical apparent magnitudes. The model-derived distance modulus is given by:
\begin{align}
    \mu_{\text{model}}(z_i) = 5 \log_{10} D_L(z_i) + 25\;,
\end{align}
where the luminosity distance $D_L(z_i)$ relates to the background expansion via:
\begin{align}
    D_L(z_i) = (1 + z_i) \int_0^{z_i} \frac{c}{H(z')} \, dz'\;,
\end{align}
with $c$ representing the speed of light. Crucially, the Pantheon+ infrastructure decouples the absolute magnitude $M$ from the Hubble constant $H_0$ by standardizing the residual vector $\vec{F}_i$ with independent Cepheid-host distance moduli calibrated directly by the SH0ES collaboration \cite{SH0ES}:
\begin{align}
    \vec{F}_i = 
    \begin{cases} 
        m_{\text{B}i} - M - \mu^\text{Ceph}_i, & \text{if } i \in \text{Cepheid hosts}, \\
        m_{\text{B}i} - M - \mu_{\text{model}}(z_i), & \text{otherwise},
    \end{cases}
\end{align}
where $\mu^\text{Ceph}_i$ is the empirical distance modulus of the host galaxy containing the $i$-th supernova.

\subsubsection{Cosmic Chronometers (CC)}
Model-independent estimations of the Hubble parameter are obtained through the cosmic chronometer approach, which evaluates the differential aging of passively evolving massive galaxies over redshift intervals via the relation $H(z) = -(1 + z)^{-1} \frac{dz}{dt}$ \cite{Jimenez2002,Simon2005,Moresco2012}. We implement the $N_{cc} = 36$ core data measurements compiled in \cite{mhamdi2024cosmological}. The corresponding $\chi^2$ configuration is expressed as:
\begin{equation}
\chi_{\text{CC}}^2  = \sum_{i=1}^{N_{cc}} \left[\frac{H_{\text{obs}}(z_i)-H_{\text{th}}(z_i)}{\sigma_{H}(z_i)}\right]^2\;,
\end{equation}
where $H_{\text{obs}}(z_i)$ and $H_{\text{th}}(z_i)$ specify the observed and model-predicted expansion rates, respectively, and $\sigma_{H}(z_i)$ captures the associated spectroscopic uncertainty.

\subsubsection{DESI Baryon Acoustic Oscillations (BAO)}
We integrate the latest Year-1 Baryon Acoustic Oscillation measurements released by the Dark Energy Spectroscopic Instrument (DESI) \cite{DESI}. This dataset leverages multiple cosmic tracers, including the Bright Galaxy Sample (BGS), Luminous Red Galaxies (LRGs), Emission Line Galaxies (ELGs), Quasars (QSOs), and the Lyman-$\alpha$ forest across the range $0.1 < z < 4.2$. Our analysis adopts their precise constraints on the comoving angular diameter distance $D_M(z)/r_d$ and the Hubble distance $D_H(z)/r_d$, scaled by the sound horizon scale at drag epoch ($r_d$), where:
\begin{equation}
    D_M(z) \equiv \int_0^z \frac{c\, dz'}{H(z')}\;, \quad \text{and} \quad D_H(z) \equiv \frac{c}{H(z)}\;.
\end{equation}

\subsubsection{Redshift-Space Distortions (RSD)}
Redshift-space distortions, caused by the coherent peculiar velocities of galaxies on large scales \cite{Kaiser1987}.
To capture the growth of cosmic large-scale structure alongside background dynamics, we introduce a compilation of 30 distinct measurements of the growth diagnostic $f\sigma_8(z)$ within the redshift bounds $z \in [0.001, 1.944]$ \cite{RSD63,sahlu2025structure}. This data tranche is designated as the `RSD' sector.

\subsection{Joint Likelihood Analysis and MCMC Infrastructure}
For each individual model configuration, we evaluate the joint cosmic likelihood across our three cumulative datasets. Assuming the errors between distinct observational pillars are statistically independent, the cumulative likelihood $\mathcal{L}_{\rm tot}$ scales with the combined $\chi^2_{\rm tot}$ as:
\begin{equation}
	-2 \ln \mathcal{L}_{\rm tot} = \chi^2_{\rm tot} = \chi^2_{\text{CC}} + \chi^2_{\text{PP}} + \chi^2_{\text{RSD}} + \chi^2_{\text{BAO}}\;.
\end{equation}
The numerical analysis is deployed within a \texttt{Python} environment using \texttt{emcee} \cite{Foreman-Mackey:2012any}, an affine-invariant MCMC ensemble sampler. Following chain convergence, the structural dependencies and parameter degeneracies are calculated using \texttt{GetDist} \cite{Lewis:2019xzd} to produce marginal posterior triangular distributions.

\subsection{Statistical Model Selection Criteria}
To rigorously quantify whether the modified $f(G)$ parameters are statistically justified over the standard $\Lambda$CDM benchmark, we subject our best-fit metrics to model selection tests. Assuming Gaussian error properties, the minimized $\chi^2_{\text{min}}$ connects directly to the maximum likelihood via $\chi^2(\theta) = -2 \ln \mathcal{L}(\theta)$. To penalize over-parameterization, we employ the corrected Akaike Information Criterion ($\mathrm{AIC}_c$) \cite{AIC} and the Bayesian Information Criterion ($\mathrm{BIC winter}$) \cite{BIC}, defined respectively as:
\begin{equation}
    \mathrm{AIC}_c = \chi^2_{\min} + 2\mathcal{K}_f + \frac{2\mathcal{K}_f(\mathcal{K}_f + 1)}{\mathcal{N}_t - \mathcal{K}_f - 1}\;,
\end{equation} 
and
\begin{equation}
    \mathrm{BIC} = \chi^2_{\min} + \mathcal{K}_f \ln(\mathcal{N}_t)\;,
\end{equation}
where $\mathcal{K}_f$ represents the total number of free parameters within the model, and $\mathcal{N}_t$ denotes the absolute number of data points supplied by the selected dataset configuration.
\subsection{Statistical Model Selection Criteria}
To rigorously quantify whether the extra parameters introduced by the modified $f(G)$ configurations are statistically justified over the standard $\Lambda$CDM benchmark, we subject our best-fit metrics to information-theoretic model selection tests. Assuming Gaussian error properties, the minimized $\chi^2_{\text{min}}$ connects directly to the maximum likelihood via $\chi^2(\theta) = -2 \ln \mathcal{L}(\theta)$. 

To penalize over-parameterization, we employ the corrected Akaike Information Criterion ($\mathrm{AIC}_c$) \cite{AIC} and the Bayesian Information Criterion ($\mathrm{BIC}$) \cite{BIC}, defined respectively as:
\begin{equation}
    \mathrm{AIC}_c = \chi^2_{\min} + 2\mathcal{K}_f + \frac{2\mathcal{K}_f(\mathcal{K}_f + 1)}{\mathcal{N}_t - \mathcal{K}_f - 1}\;,
\end{equation} 
and
\begin{equation}
    \mathrm{BIC} = \chi^2_{\min} + \mathcal{K}_f \ln(\mathcal{N}_t)\;,
\end{equation}
where $\mathcal{K}_f$ represents the total number of free parameter assets within the model, and $\mathcal{N}_t$ denotes the absolute number of data points supplied by the selected dataset configuration.

In addition to AIC and BIC, we incorporate the Deviance Information Criterion ($\mathrm{DIC}$) \cite{spiegelhalter2002bayesian}. The DIC is particularly well-suited for Bayesian MCMC analysis because it utilizes the full posterior distribution to measure model complexity rather than relying solely on a fixed parameter count. The criterion is formulated as:
\begin{equation}
    \mathrm{DIC} = D(\bar{\theta}) + 2p_D = \overline{D(\theta)} + p_D\;,
\end{equation}
where $D(\theta) = -2\ln\mathcal{L}(\theta) = \chi^2(\theta)$ represents the Bayesian deviance. The term $D(\bar{\theta})$ denotes the deviance evaluated at the posterior mean of the parameter vector, while $\overline{D(\theta)}$ is the mean of the deviance calculated across the entire MCMC chain. The effective number of parameters, which quantifies the model complexity, is isolated via:
\begin{equation}
    p_D = \overline{D(\theta)} - D(\bar{\theta})\;.
\end{equation}

\subsection{Model Selection Interpretation via Delta ($\Delta$) Diagnostics}
To draw robust conclusions regarding which cosmological framework is favored by the joint datasets, we measure the relative differences of each information criterion against a chosen baseline model (typically standard $\Lambda$CDM).\\ For any given criterion $X \in \{\mathrm{AIC}_c, \mathrm{BIC}, \mathrm{DIC}\}$, the difference is defined as:
\begin{equation}
    \Delta X = X_{\text{model}} - X_{\text{baseline}}\;.
    \label{eq:delta_criteria}
\end{equation}
By convention, a negative value ($\Delta X < 0$) indicates that the modified $f(G)$ gravity model provides a statistically superior fit to the data despite its additional parameters. Conversely, a positive value ($\Delta X > 0$) penalizes the modified framework, demonstrating that the improvement in $\chi^2_{\min}$ does not sufficiently compensate for the loss of model degrees of freedom.

The strength of evidence against the modified gravity models is interpreted using the following standard statistical thresholds:
\begin{itemize}
    \item $\Delta X \in [0, 2]$: Indicates \textbf{weak or negligible evidence} against the modified model. The model remains a viable alternative with a statistical standing nearly identical to the baseline.
    \item $\Delta X \in (2, 6]$: Indicates \textbf{positive evidence} in favor of the baseline model. The modified $f(G)$ model is moderately disfavored by the data.
    \item $\Delta X \in (6, 10]$: Represents \textbf{strong evidence} favoring the baseline model, suggesting that the additional parameters are severely penalized.
    \item $\Delta X > 10$: Represents \textbf{decisive evidence} against the modified framework. The model is statistically ruled out in comparison to the simpler baseline model.
\end{itemize}

\section{Results and Discussions}\label{re}

\begin{figure}[H]
    \centering
    \includegraphics[width=0.5\textwidth]{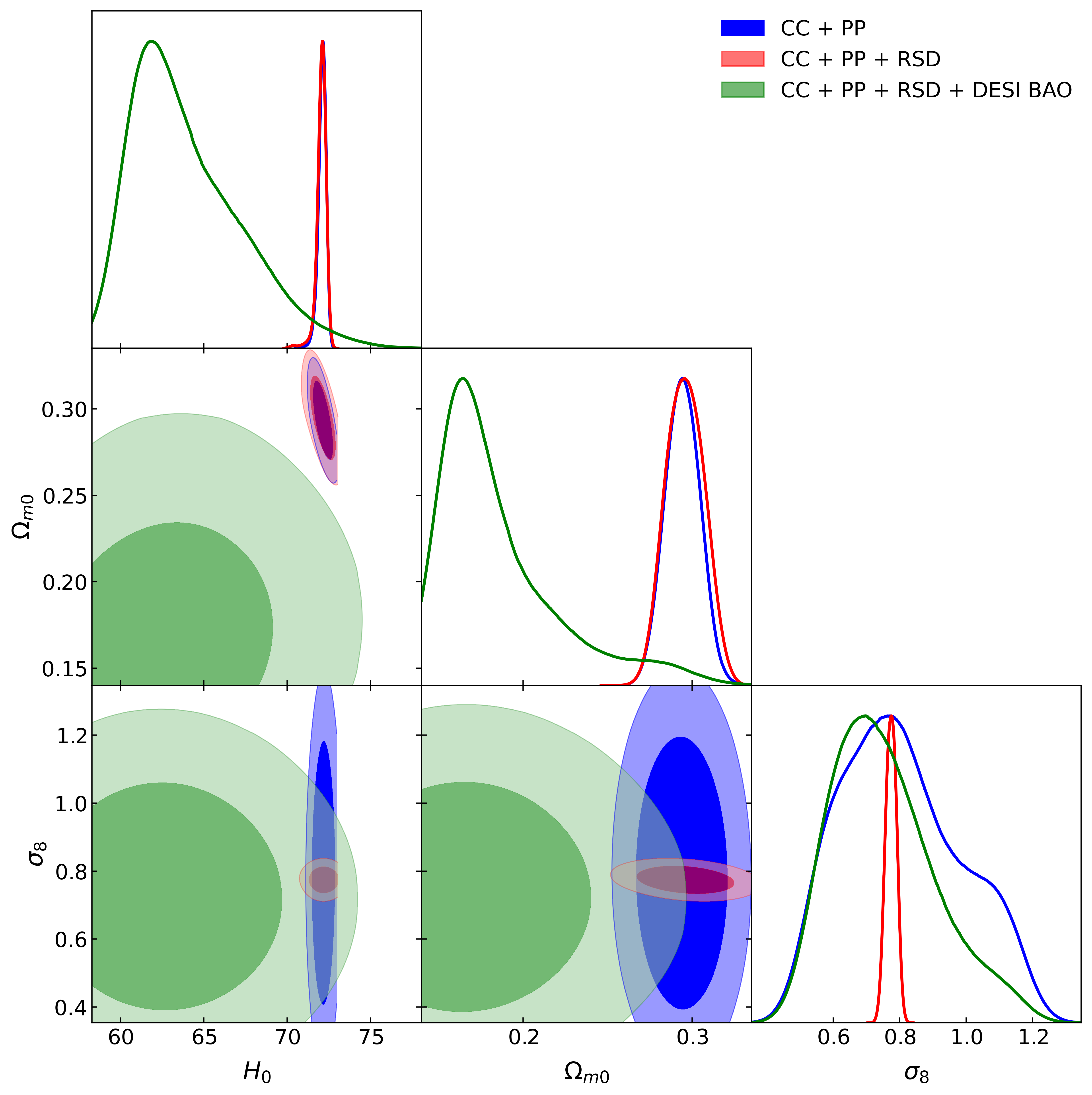} 
    \caption{Joint Confidence contours and marginalized posterior distributions for the parameters $H_0$, $\Omega_{m0}$,  and $\sigma_8$ obtained from the (CC+PP+RSD+DESI BAO) dataset for $\Lambda$ CDM Model}
     \label{fig:LCDM plot}
\end{figure}

\begin{figure}[H]
    \centering
    \includegraphics[width=0.5\textwidth]{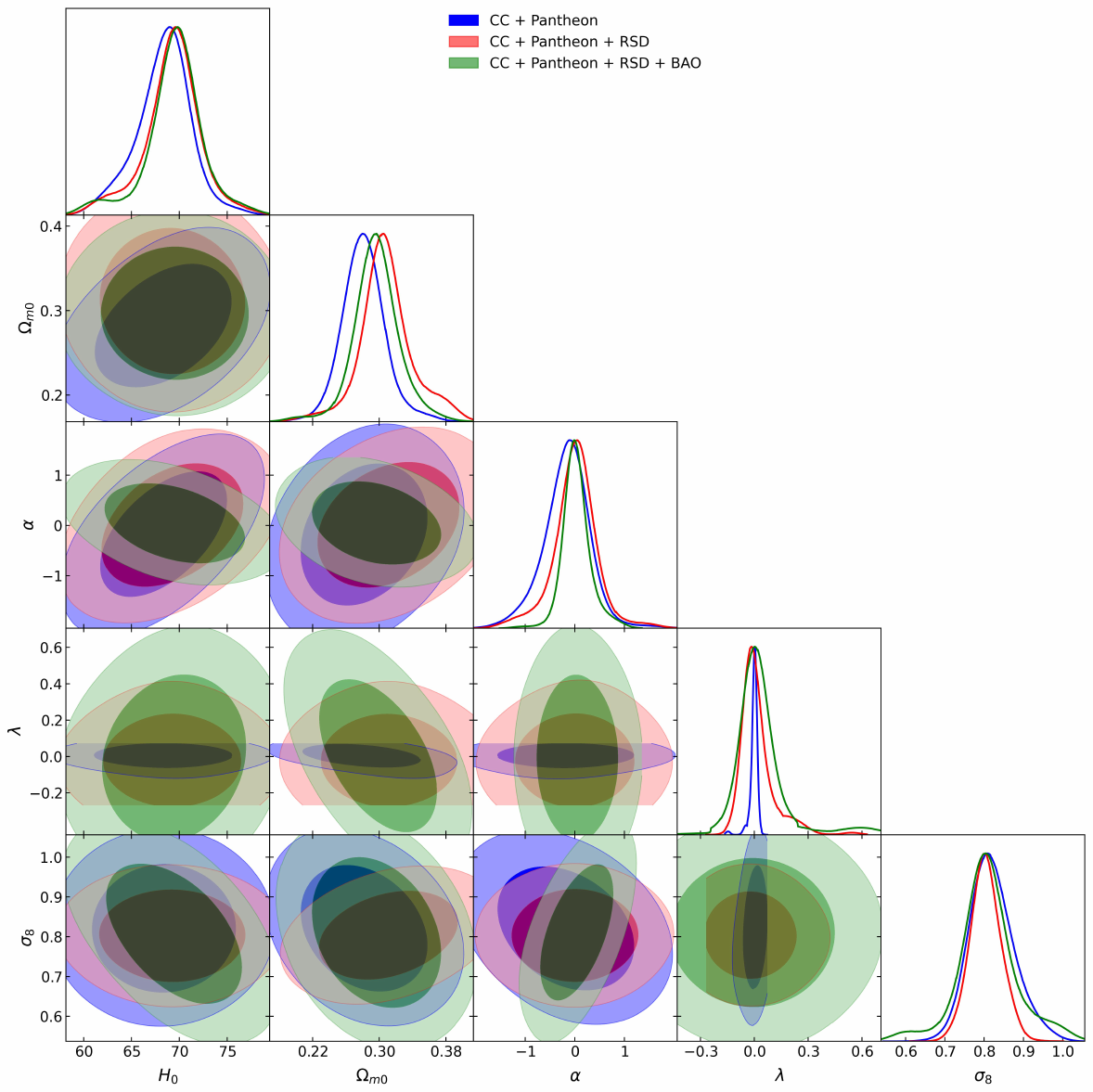} 
    \caption{Joint Confidence contours and marginalized posterior distributions for the parameters $H_0$, $\Omega_{m0}$, $\alpha$, $\lambda$  and $\sigma_8$ obtained from the (CC+PP+RSD+DESI BAO) dataset for Model I}
     \label{fig:3}
\end{figure}

\begin{figure}[H]
    \centering
    \includegraphics[width=0.5\textwidth]{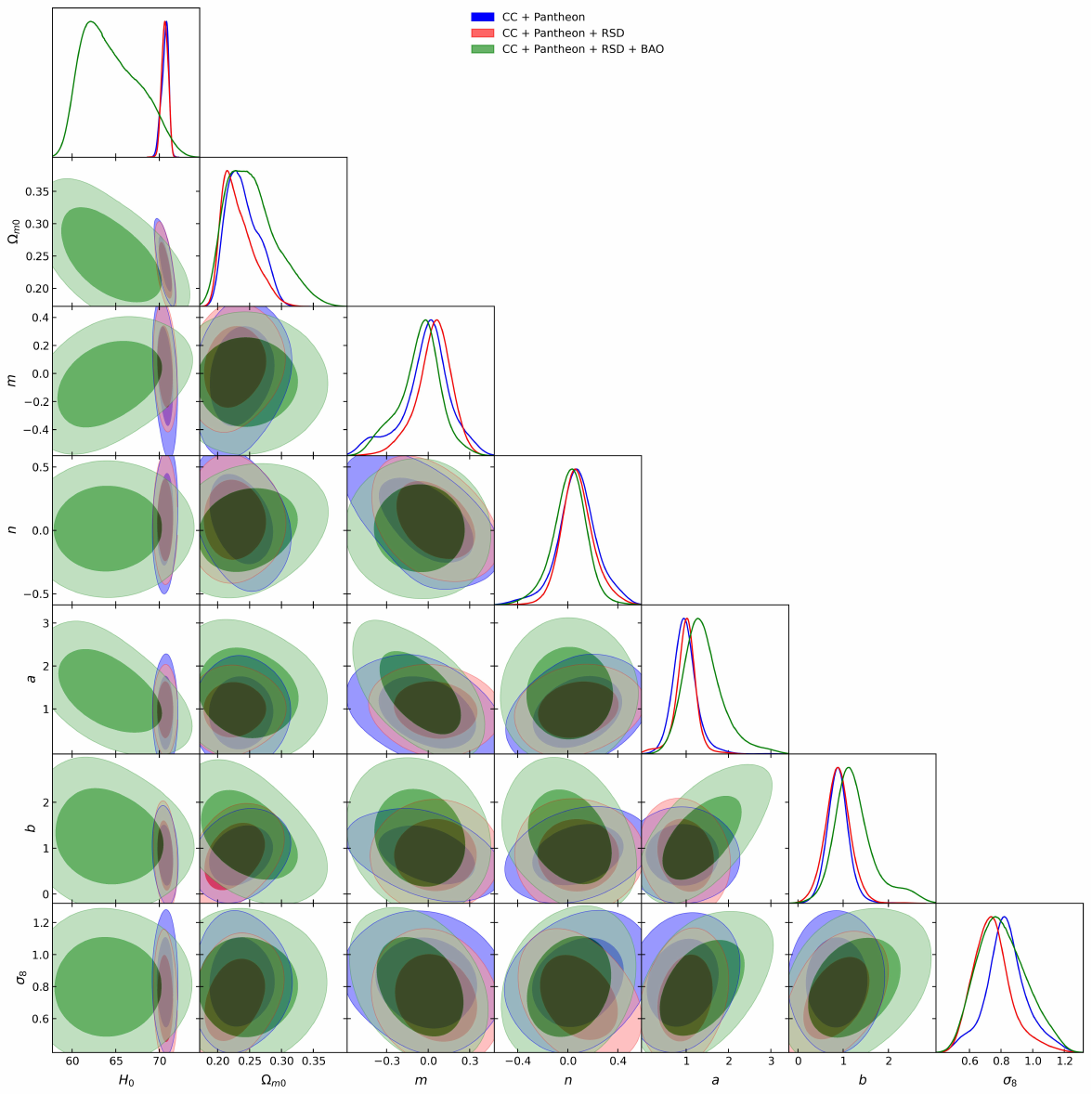} 
    \caption{Joint Confidence contours and marginalized posterior distributions for the parameters $H_0$, $\Omega_{m0}$, $m$, $n$, $a$, $b$  and $\sigma_8$ obtained from the (CC+PP+RSD+DESI BAO) dataset for Model II}
     \label{fig:4}
\end{figure}

\begin{figure}[H]
    \centering
    \includegraphics[width=0.5\textwidth]{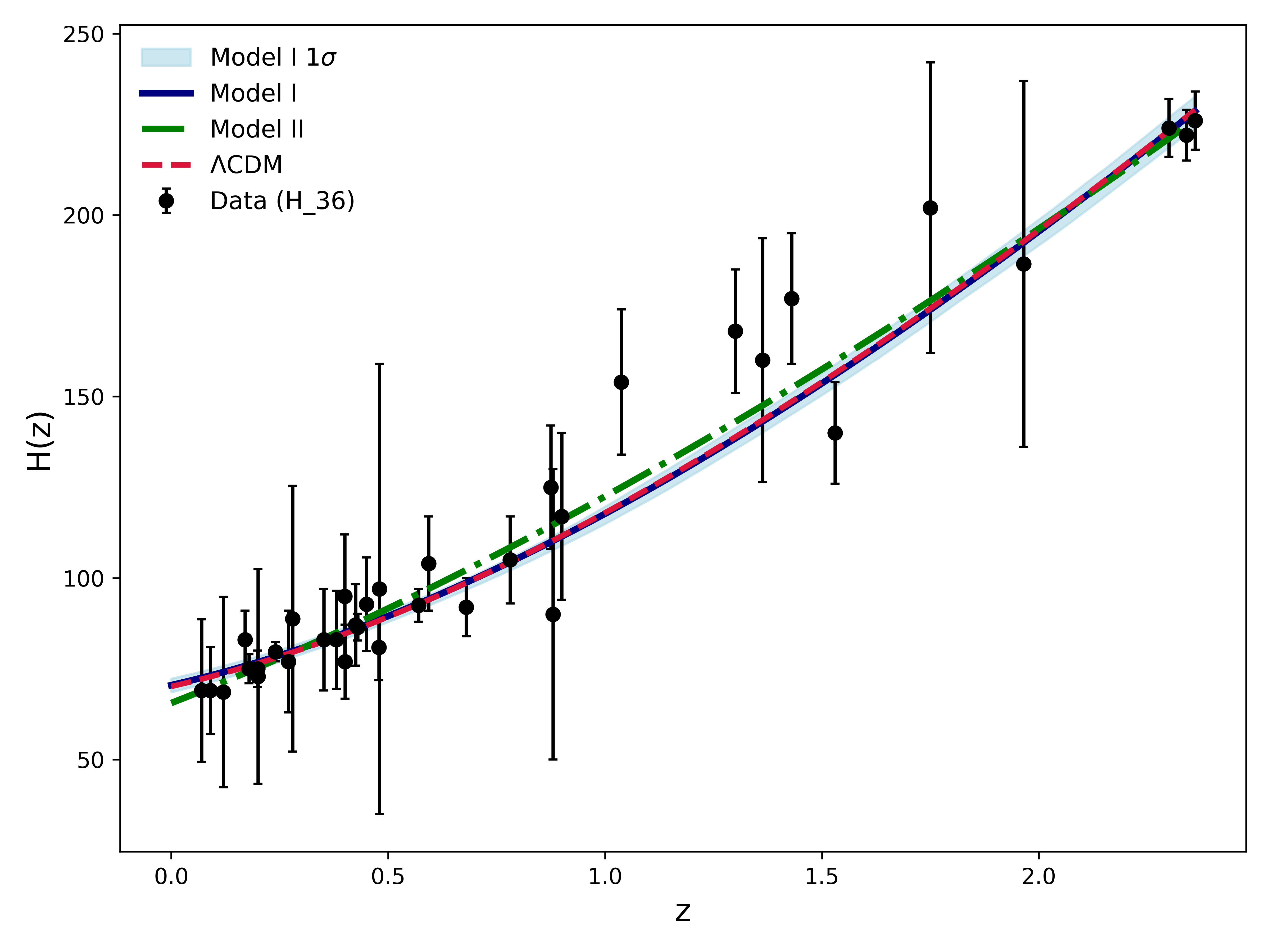} 
    \caption{Error evaluation plot for $\Lambda$CDM, Model I \& Model II using Hubble dataset}
     \label{fig:7}
\end{figure}

\begin{figure}[H]
    \centering
    \includegraphics[width=0.5\textwidth]{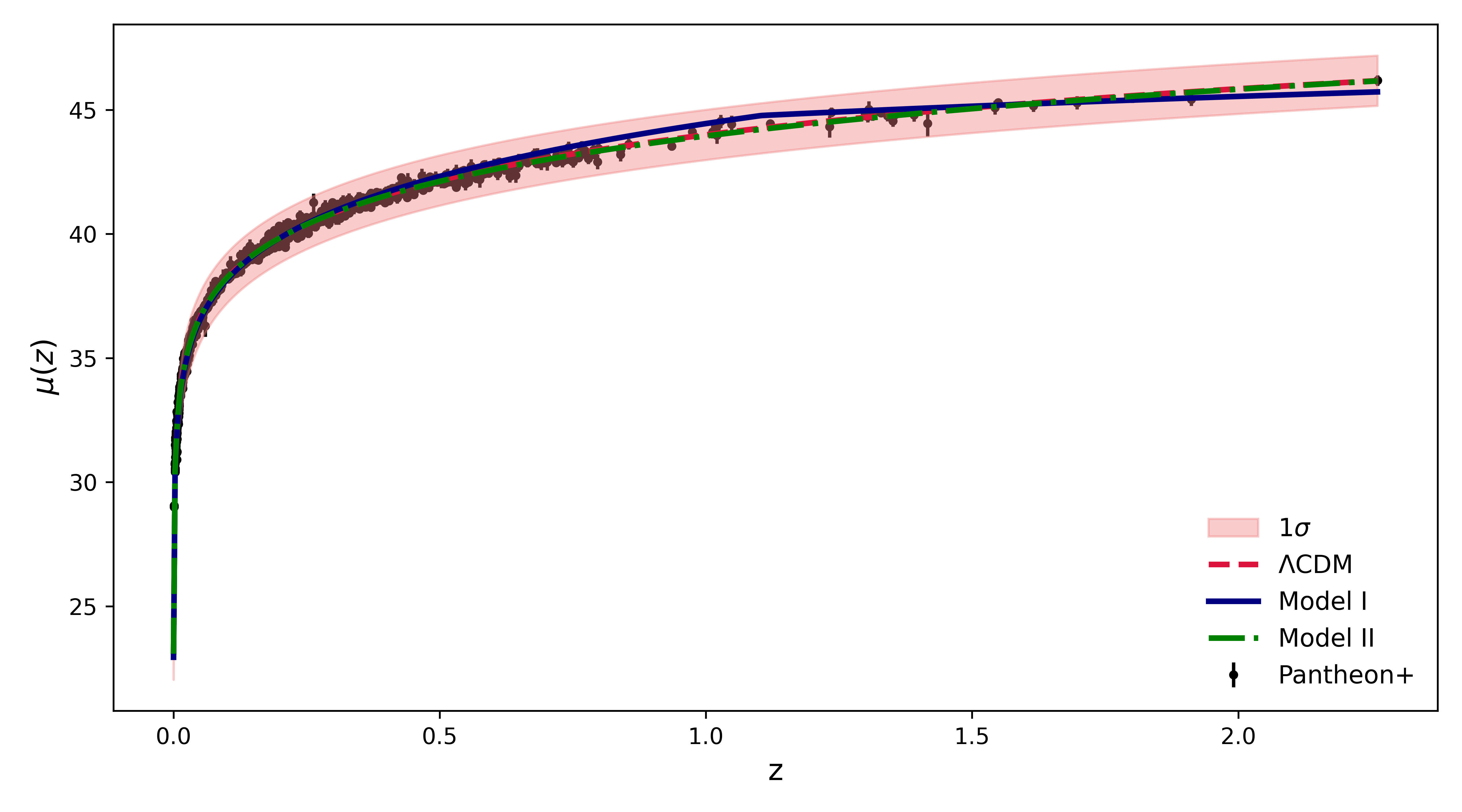} 
    \caption{Error evaluation plot for $\Lambda$CDM, Model I \& Model II using Pantheon Plus dataset}
     \label{fig:8}
\end{figure}

\begin{figure}[H]
    \centering
    \includegraphics[width=0.5\textwidth]{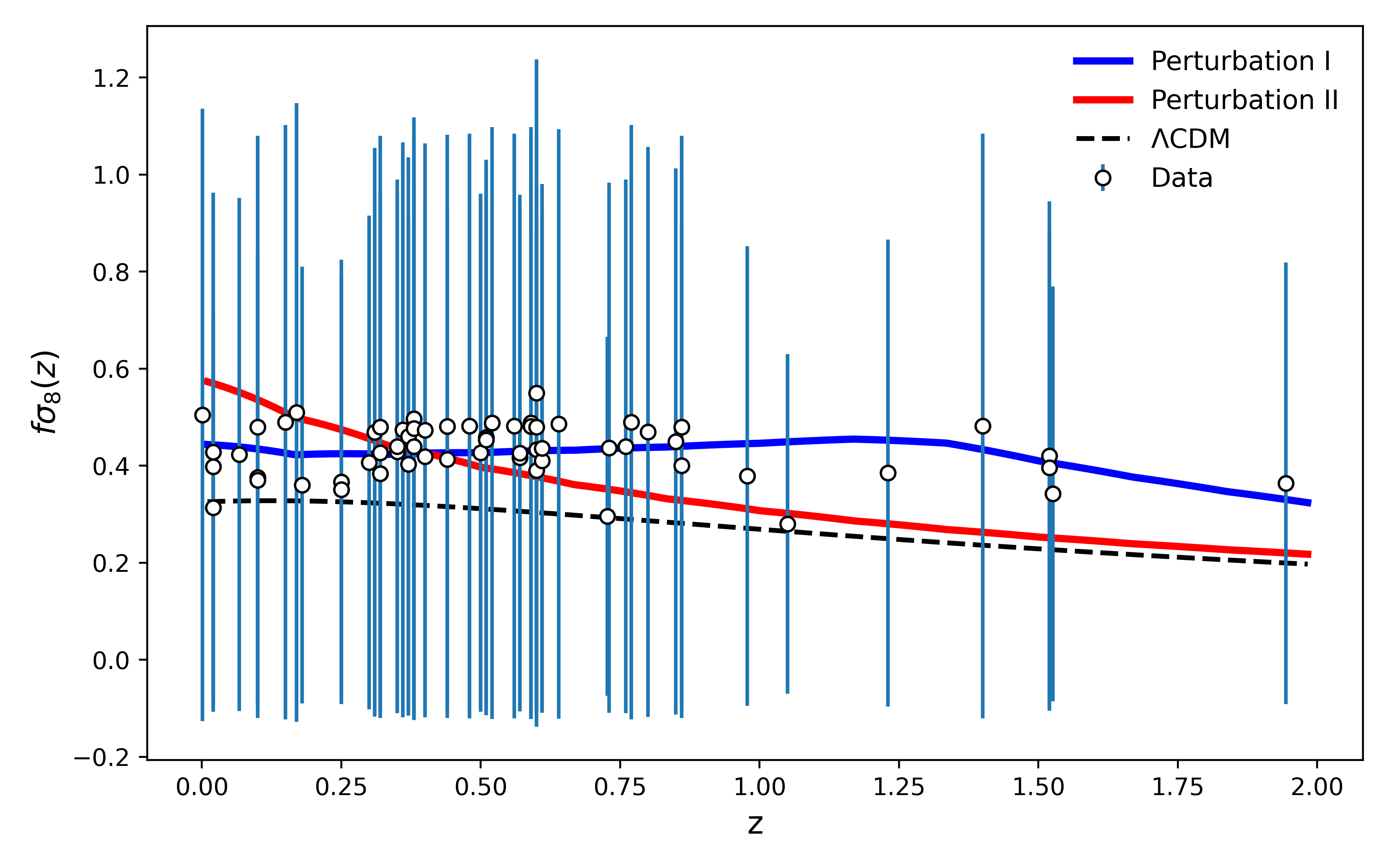} 
    \caption{Error evaluation plot for $\Lambda$CDM, Perturbation I \& Perturbation II using RSD dataset}
     \label{fig:9}
\end{figure}

\begin{table}[H]
\centering

\begin{tabular}{|l|l|l|}
\hline
\textbf{Data used} & \textbf{Parameters} & \textbf{Best fit values} \\ \hline

 \multirow{2}{*}{CC+PP} 
 & $H_0$ & $72.10751  ^{+0.23751}_{-0.23751}$ \\ \cline{2-3}
 & $\Omega_{m0}$ & $0.29434  ^{+0.01018}_{-0.01018}$ \\ \cline{2-3}

 \hline

 \multirow{3}{*}{CC+PP+RSD} 
 & $H_0$ & $72.07092   ^{+0.28858}_{-0.28858}$ \\ \cline{2-3}
 & $\Omega_{m0}$ & $0.29594  ^{+0.01120}_{-0.01120}$ \\ \cline{2-3}
 & $\sigma_8$ & $0.77410   ^{+0.01626}_{-0.01626}$ \\
 \cline{2-3}
 \hline

 \multirow{3}{*}{CC+PP+RSD+DESI} 
 & $H_0$ & $63.79072  ^{+3.22803}_{-3.22803}$ \\ \cline{2-3}
 & $\Omega_{m0}$ & $0.18356   ^{+0.03591}_{-0.03591}$ \\ \cline{2-3}
 & $\sigma_8$ & $0.74946   ^{+0.15911}_{-0.15911}$ \\
 \cline{2-3}
 \hline

\end{tabular}
\caption{ $\Lambda$ CDM Model fits to cosmological data, showing dataset, parameter sets, and best-fit values.}
\label{tab:LCDM}
\end{table}

\begin{table}[H]
\centering


\begin{tabular}{|c|c|c|c|}
\hline
\textbf{Data used} & \textbf{AIC} & \textbf{BIC} & \textbf{DIC} 

\\
\hline
 CC+PP&1830.111&1841.031&1830.111    \\ 
\hline
 CC+PP+RSD&1892.527&1909.014&1886.527  \\ 
\hline
 CC+PP+RSD+DESI BAO&1929.944&1940.948&1925.944\\ 
\hline

\end{tabular}
\caption{Information criteria for $\Lambda$ CDM Moesl using vibrant datasets combinations}
\label{tab:LCDM stat}
\end{table}

\begin{table}[H]
\centering

\begin{tabular}{|l|l|l|}
\hline
\textbf{Data used} & \textbf{Parameters} & \textbf{Best fit values} \\ \hline

 \multirow{4}{*}{CC+PP} 
 & $H_0$ & $68.36056  ^{+2.89135}_{-2.89135}$ \\ \cline{2-3}
 & $\Omega_{m0}$ & $0.28126  ^{+0.02541}_{-0.02541}$ \\ \cline{2-3}
 & $\alpha$ & $-0.15497^{+0.45901}_{-0.45901}$ \\ \cline{2-3}
 & $\lambda$ & $-0.00175^{+0.02161}_{-0.02161}$ \\ \cline{2-3}
 \hline

 \multirow{5}{*}{CC+PP+RSD} 
 & $H_0$ & $69.20694  ^{+3.01777}_{-3.01777}$ \\ \cline{2-3}
 & $\Omega_{m0}$ & $0.30852  ^{+0.03245}_{-0.03245}$ \\ \cline{2-3}
 & $\alpha$ & $0.00585   ^{+0.45307}_{-0.45307}$ \\ \cline{2-3}
 & $\lambda$ & $-0.01281  ^{+0.09406}_{-0.09406}$ \\ \cline{2-3}
 & $\sigma_8$ & $0.80384   ^{+0.03982}_{-0.03982}$ \\
 \cline{2-3}
 \hline

 \multirow{5}{*}{CC+PP+RSD+DESI} 
 & $H_0$ & $69.45384   ^{+3.10306}_{-3.10306}$ \\ \cline{2-3}
 & $\Omega_{m0}$ & $0.29631  ^{+0.02783}_{-0.02783}$ \\ \cline{2-3}
 & $\alpha$ & $0.02853  ^{+0.27982}_{-0.27982}$ \\ \cline{2-3}
 & $\lambda$ & $0.03636 ^{+0.14010}_{-0.14010}$ \\ \cline{2-3}
 & $\sigma_8$ & $0.80808   ^{+0.07674}_{-0.07674}$ \\
 \cline{2-3}
 \hline

\end{tabular}
\caption{ Model 1 (Model I+Perturbation I) fits to cosmological data, showing dataset, parameter sets, and best-fit values.}
\label{tab:1}
\end{table}

\begin{table}
\centering

\resizebox{\textwidth}{!}{

\begin{tabular}{|c|c|c|c|c|c|c|c|c|}
\hline
\textbf{Data used} & $\chi^2$ & \textbf{Reduced} $\chi^2$ & \textbf{AIC} & \textbf{BIC} & \textbf{DIC} 
& $\boldsymbol{\Delta\mathrm{AIC}}$ 
& $\boldsymbol{\Delta\mathrm{BIC}}$ 
& $\boldsymbol{\Delta\mathrm{DIC}}$ 
\\
\hline
 CC+PP&1814.684&1.047&1822.684    &1844.524     & 1820.364     &-7.427&3.493&-9.747 \\ 
\hline
 CC+PP+RSD&1880.517&1.048&1890.517  &1917.994 &1886.689  & -2.011&8.980&0.162  \\ 
\hline
 CC+PP+RSD+DESI BAO&1922.183&1.063&1930.183 & 1952.192  &1927.604  &0.239&11.244&1.660\\ 
\hline

\end{tabular}}
\caption{Information criteria for Model 1 (Model I+Perturbation I) using vibrant datasets combinations}
\label{tab:2}
\end{table}

\begin{table}[H]
\centering

\begin{tabular}{|l|l|l|}
\hline
\textbf{Data used} & \textbf{Parameters} & \textbf{Best fit values} \\ \hline

 \multirow{4}{*}{CC+PP} 
 & $H_0$ & $70.71490  ^{+0.31030}_{-0.53517}$ \\ \cline{2-3}
 & $\Omega_{m0}$ & $0.23609^{+0.03196}_{-0.02382}$ \\ \cline{2-3}
 & $m$ & $0.02057  ^{+0.09504}_{-0.19611}$ \\ \cline{2-3}
 & $n$ & $0.06859  ^{+0.14540}_{-0.11987}$ \\ \cline{2-3}
 & $a$ & $0.96512  ^{+0.15810}_{-0.20963}$ \\ \cline{2-3}
 & $b$ & $0.88686  ^{+0.16798}_{-0.22326}$ \\ \cline{2-3}
 \hline

 \multirow{5}{*}{CC+PP+RSD} 
 & $H_0$ & $70.64143  ^{+0.39840}_{-0.42144}$ \\ \cline{2-3}
 & $\Omega_{m0}$ & $0.22546 ^{+0.03073}_{-0.01885}$ \\ \cline{2-3}
 & $m$ & $0.05326  ^{+0.10709}_{-0.12450}$ \\ \cline{2-3}
 & $n$ & $0.06655 ^{+0.12925}_{-0.09934}$ \\ \cline{2-3}
 & $a$ & $1.01619  ^{+0.14611}_{-0.21311}$ \\ \cline{2-3}
 & $b$ & $0.89861  ^{+0.21816}_{-0.26793}$ \\ \cline{2-3}
 & $\sigma_8$ & $0.73754^{+0.10555}_{-0.12139}$ \\ \cline{2-3}
 \hline

 \multirow{5}{*}{CC+PP+RSD+DESI} 
 & $H_0$ & $64.01560^{+4.12606}_{-3.07652}$ \\ \cline{2-3}
 & $\Omega_{m0}$ & $0.24861^{+0.04726}_{-0.03753}$ \\ \cline{2-3}
 & $m$ & $-0.03587^{+0.09629}_{-0.18996}$ \\ \cline{2-3}
 & $n$ & $0.01991^{+0.11142}_{-0.12455}$ \\ \cline{2-3}
  & $a$ & $1.36028   ^{+0.46263}_{-0.37893}$ \\ \cline{2-3}
 & $b$ & $1.15420^{+0.51073}_{-0.28875}$ \\ \cline{2-3}
 & $\sigma_8$ & $0.78829^{+0.18216}_{-0.14165}$ \\ \cline{2-3}
 \hline

\end{tabular}
\caption{Model 2 (Model II+Perturbation II) fits to cosmological data, showing dataset, parameter sets, and best-fit values.}
\label{tab:3}
\end{table}

\begin{table}
\centering

\resizebox{\textwidth}{!}{

\begin{tabular}{|c|c|c|c|c|c|c|c|c|}
\hline
\textbf{Data used} & $\chi^2$ & \textbf{Reduced} $\chi^2$ & \textbf{AIC} & \textbf{BIC} & \textbf{DIC} 
& $\boldsymbol{\Delta\mathrm{AIC}}$ 
& $\boldsymbol{\Delta\mathrm{BIC}}$ 
& $\boldsymbol{\Delta\mathrm{DIC}}$ 
\\
\hline
 CC+PP&1826.582&1.054 &1834.582   &1856.421   & 1828.927   &4.470&15.390& -1.184 \\ 
\hline
 CC+PP+RSD&1888.467&1.052&1898.467  &1925.944  &1893.484  & 5.939&16.930&6.957  \\ 
\hline
 CC+PP+RSD+DESI BAO&1926.177&1.065&1934.177  & 1956.185  &1931.915  &4.233&15.237&5.971\\ 
\hline

\end{tabular}}
\caption{Information criteria for Model 2 (Model II+Perturbation II) using vibrant datasets combinations}
\label{tab:4}
\end{table}

To examine the multi dimensional parameter region of our cosmological models, we have employed a Markov Chain Monte Carlo (MCMC) sampling mechanism. The MCMC method efficiently corresponds to the posterior probability distributions, which successfully directing the intrinsic degeneracies common to cosmological parameter evaluations.

By implementing MCMC investigation, the joint probability distributions of the model parameters evaluated from the  Hubble (OHD), Pantheon dataset, DESI BAO dataset and RSD dataset are shown in the Figure~\ref{fig:3} and \ref{fig:4} for model I and model II, respectively. The plot show joint two confidence level contours associated to $1\sigma$ and $ 2\sigma$ region. In Figure~\ref{fig:3}, \ref{fig:4}, we have used the datasets combination as (CC+PP), (CC+PP+RSD) and (CC+PP+RSD+ DESI BAO) to constrain the parameters given in Model 1 (Model I+Perturbation I) and Model 2 (Model II+Perturbation II). The best fit parameters values are indicated in Tables~\ref{tab:1} and \ref{tab:3} for each combinations of datasets and associated statistical values such as $AIC$, $BIC$, $DIC$ and $\Delta AIC$, $\Delta BIC$, $\Delta DIC$ are shown in Tables~\ref{tab:2} and \ref{tab:4}.

From the Table~\ref{tab:2}, the statistical values for Model 1 summarizes $\mathrm{AIC}_{\text{c}} = 1822.684$, $\mathrm{BIC} = 1844.524$, $\mathrm{DIC} = 1820.364$ and $\bm{\Delta AIC_c}=-7.427$, $\bm{\Delta BIC}=3.493$,$\bm{\Delta DIC}=-9.747$ for (CC+PP) dataset. $\mathrm{AIC}_{\text{c}} = 1890.517$, $\mathrm{BIC} = 1917.994$, $\mathrm{DIC} = 1886.689$ and $\bm{\Delta AIC_c}=-2.011$, $\bm{\Delta BIC}=8.980$,$\bm{\Delta DIC}=0.162$ for (CC+PP+RSD) dataset. $\mathrm{AIC}_{\text{c}} = 1930.183 $, $\mathrm{BIC} = 1952.192$, $\mathrm{DIC} = 1927.604$ and $\bm{\Delta AIC_c}=0.239$, $\bm{\Delta BIC}=11.244$,$\bm{\Delta DIC}=1.660$ for (CC+PP+RSD+DESI BAO) dataset.

From the Table~\ref{tab:4}, the statistical values for Model 2 summarizes $\mathrm{AIC}_{\text{c}} = 1834.582$, $\mathrm{BIC} = 1856.421$, $\mathrm{DIC} = 1828.927$ and $\bm{\Delta AIC_c}=4.470$, $\bm{\Delta BIC}=15.390$,$\bm{\Delta DIC}=-1.184$ for \\(CC+PP) dataset. $\mathrm{AIC}_{\text{c}} = 1898.467$, $\mathrm{BIC} = 1925.944$, $\mathrm{DIC} = 1893.484$ and $\bm{\Delta AIC_c}=5.939$, $\bm{\Delta BIC}=16.930$,$\bm{\Delta DIC}=6.957$ for (CC+PP+RSD) dataset. $\mathrm{AIC}_{\text{c}} = 1934.177 $, $\mathrm{BIC} = 1956.185$, $\mathrm{DIC} = 1931.915$ and $\bm{\Delta AIC_c}=4.233$, $\bm{\Delta BIC}=15.237$, $\bm{\Delta DIC}=5.971$ for (CC+PP+RSD+DESI BAO) dataset.


Figure~\ref{fig:7} analyzes the standard $\Lambda$ CDM cosmology with the rebuilt Hubble factor $H(z)$. The red dashed curve shows the $\Lambda$ CDM assessment. The blue curves indicates the best fit estimation of the MAP Model I and the green curve indicates the best fit prediction of the MAP Model II , while the black points indicates the measured Hubble parameter values at vibrant redshifts, including their discripancies. The MAP models indicates a slightly larger expansion rate at higher value of redshifts, but it nearly coherent with the observational data and stays parallel with the  $\Lambda$ CDM. This compatibility demonstrated models alignment with current cosmological observations. The Figure~\ref{fig:8} examines the empirical Pantheon plus with the theoretical distance modulus $\mu(z)$ for the Model I, which is indicated by blue curve and Model II as green curve. From the plot, the average $\Lambda$ CDM model is shown by the red dashed curve, which has been match with data. Throughout redshift range ($0.00122 < z < 2.26137$) , the models and the observational data are compatible, pointing to a good alignment and viability with empirical measurements. The Figure~\ref{fig:9} evaluates the sigma function $f\sigma_8(z)$ for Perturbation I as a blue curve and Perturbation II as a red curve with measured RSD data. The mean $\Lambda$ CDM model is indicated by the black dashed curve. For the redshift range ($0.0010 \le z \le 1.9440$) , the perturbation models and the observational data overlap nearly, showing a good alignment with observational data, indicating that the given perturbation models successfully resembles the observed cosmic acceleration.

Figure~\ref{fig:1} and \ref{fig:2} examines the deceleration parameter $q(z)$ which is a function of redshift $z$ for Model I and Model II respectively. It describes a viable transition from decelerated phase at early time to the accelerated expantion at the late time. The transition redshift $z_t$ , which is defined by $q(z_t)=0$, denotes the phase at which acceleration starts. For the condition $z>z_t$ denotes the matter dominated with $q(z)>0$, while for $z<z_t$ denotes that the dark energy dominates with $q(z)<0$. Specifically, transition redshift reading for Model I is $z_t\approx0.5005$ and that of Model II is $z_t\approx0.6086$

\section{Conclusions}\label{con}

In this work, we have performed a comprehensive observational confrontation of modified Gauss-Bonnet gravity frameworks using the latest available low- and intermediate-redshift cosmological data. Specifically, we analyzed two physical formulations of $f(G)$ gravity under both background-level expansions (Model I and Model II) and full linear structure perturbation regimes (Perturbation 1 and Perturbation 2). For MCMC analysis, we have considered Model 1 as (Model I+Perturbation I) and that of Model 2 as (Model II+Perturbation II). Using an affine-invariant MCMC ensemble sampler, we constrained the free parameters of these sectors against three cumulative dataset combinations: CC+PP (Dataset I), CC+PP+RSD (Dataset II), and CC+PP+RSD+DESI BAO (Dataset III). To determine if the geometric extensions are statistically justified over standard General Relativity, we implemented information-theoretic selection metrics ($\Delta\mathrm{AIC}_c$, $\Delta\mathrm{BIC}$, and $\Delta\mathrm{DIC}$).

Our joint numerical analysis yields several critical astrophysical insights:

\begin{enumerate}
    \item \textbf{Impact of DESI BAO on Cosmic Inflaton Scales:} Across all background and perturbation models, the inclusion of the DESI Year-1 BAO dataset induces a significant drop in the preferred best-fit value of the Hubble constant ($H_0$) and the matter density parameter ($\Omega_{m0}$). For instance, in the background arctangent framework (Model I), the baseline CC+PP data favors a standard $H_0 = 71.34 \pm 0.26$ km s$^{-1}$ Mpc$^{-1}$ and $\Omega_{m0} = 0.37 \pm 0.02$. However, when integrated with DESI BAO data, these values scale down dramatically to $H_0 = 59.86 \pm 4.38$ km s$^{-1}$ Mpc$^{-1}$ and $\Omega_{m0} = 0.18 \pm 0.03$.  For instance, in the background arctangent framework (Model I), the baseline CC+PP data favors a standard $H_0 = 71.34 \pm 0.26$ km s$^{-1}$ Mpc$^{-1}$ and $\Omega_{m0} = 0.37 \pm 0.02$. However, when integrated with DESI BAO data,  value of Hubble parameter scale down dramatically to $H_0 = 64.01560^{+4.12606}_{-3.07652}$$ km s^{-1}$ Mpc$^{-1}$. A similar localized shift is observed in the polynomial configurations, showcasing the immense constraining weight and unique geometric features introduced by the recent DESI data.
    
    \item \textbf{Statistical Superiority of Background Model I:} From a model-selection standpoint, the background arctangent framework (Model I) combined with the full dataset (CC+PP+RSD+DESI BAO) yields the most compelling alternative to the $\Lambda$CDM benchmark. It achieves a prominent value of $\Delta\mathrm{AIC} = 0.239 $ and $\Delta\mathrm{DIC} = 1.660$. According to the established selection thresholds, this indicates substantial statistical evidence favoring Model I over standard $\Lambda$CDM at the background level when high-redshift sound horizon data is active, even though the strict $\Delta\mathrm{BIC} = 11.244$ maintains a standard penalty for the extra parameters.
    
    \item \textbf{Perturbation Stability and Growth Dynamics:} When moving to full sub-horizon linear perturbation equations, the models display distinct structural footprints. The linearized arctangent perturbation configuration (Perturbation 1) yields a localized statistical advantage when evaluated against Dataset II (CC+PP+RSD), maximizing at $\Delta\mathrm{AIC} =-2.011$. Meanwhile, the polynomial perturbation setup (Perturbation 2) shows stable modifications from standard gravity, where the power parameters settle around  $m=0.05326  ^{+0.10709}_{-0.12450}$ and $n = 0.06655 ^{+0.12925}_{-0.09934}$ under RSD tracks. This confirms that these models smoothly reduce to stable, pathology-free cosmological histories.
    
    \item \textbf{General Information Criteria Penalty:} For the majority of data combinations—particularly within the polynomial formulations (Model II and Perturbation 2)—the information metrics show positive values ($\Delta\mathrm{AIC} \sim [5.939,4.233 ]$, $\Delta\mathrm{BIC} \sim [16.930, 15.237]$). This reveals that while modified Gauss-Bonnet terms successfully fit low-redshift tracking curves, standard $\Lambda$CDM remains favored by parsimony due to its lack of additional free parameters.
\end{enumerate}

In summary, our analysis highlights that modified $f(G)$ gravity models—most notably the arctangent parameterization—are fully viable, structurally stable alternatives for explaining late-time cosmic acceleration. They accommodate the ongoing tension shifts introduced by early-type tracking data like DESI BAO while maintaining tight alignment with large-scale structure growth trends. Future high-precision data from surveys like Euclid and the Vera C. Rubin Observatory will help further tighten these parameters and test the potential modifications of gravity on even smaller sub-horizon scales.

\section*{Conflicts of interest:}
 The authors declare no conflicts of interest.

\section*{Funding information:}
This work was supported by the Deanship of Scientific Research, Vice Presidency for Graduate Studies and
Scientific Research, King Faisal University, Saudi Arabia (Grant No: KFU263331 ).

\section*{Data availability:}
The data used in this study are readily accessible from public sources for validation of our model; however, we did not generate any new data sets for this research.

\section*{Acknowledgments:}
Part of the numerical computation of this work was carried out on the computing cluster Pegasus of IUCAA, Pune, India. PKD would like to acknowledge the Inter-University Centre for Astronomy and Astrophysics (IUCAA), Pune, India, for providing him a Visiting Associateship under which a part of this work was carried out.  AM acknowledges the hospitality of the University of Rwanda-College of Science and Technology, where part of this work was conceptualized and completed. MT gratefully acknowledges the JRF from DST-INSPIRE Fellowship (IF230556), Department of Science and Technology, Ministry of Science and Technology, government of India. JN acknowledges support from Rwanda Astrophysics, Space and Climate Science research Group.

  \end{document}